\documentclass{WileyMSP-template}
\usepackage{diagbox}
\usepackage{textcomp}
\usepackage{textgreek}
\usepackage{cite}
\usepackage{hyperref}
\usepackage{academicons}
\usepackage{xcolor}
\usepackage{adjustbox}
\usepackage{multirow}
\usepackage{amsmath}
\usepackage{amssymb}
\usepackage[version=4]{mhchem}
\usepackage[labelfont=bf,labelsep=colon]{caption}
\usepackage{xcolor}
\definecolor{linkcolor}{rgb}{0.5,0.1,0.1}
\definecolor{urlcolor} {rgb}{0.1,0.1,0.5}
\definecolor{citecolor}{rgb}{0.1,0.5,0.1}
\hypersetup{colorlinks=true,linkcolor=linkcolor,urlcolor=urlcolor,citecolor=citecolor,pdfborder={0 0 0}}

\newcommand{\orcid}[1]{\href{https://orcid.org/#1}{\textcolor[HTML]{A6CE39}{\aiOrcid}}}

\title{\textit{Anoplophora graafi} Longhorn Beetle Coloration is due to Disordered Diamond-like Packed Spheres}

\begin{document}
\maketitle

\author{Kenza Djeghdi}
\author{C\'edric Schumacher}
\author{Ilja Gunkel}
\author{Bodo D.\ Wilts*}
\author{Ullrich Steiner*}

\begin{affiliations}
K.\ Djeghdi, C.\ Schumacher, I.\ Gunkel, U.\ Steiner\\
Adolphe Merkle Institute and National Competence Center in Bioinspired Materials, University of Fribourg, Chemin des Verdiers 4, 1700 Fribourg, Switzerland\\
Email Address: ullrich.steiner@unifr.ch

B.D. Wilts\\
Department of Chemistry and Physics of Materials, University of Salzburg, Jakob-Haringer-Straße 2A, 5020 Salzburg, Austria and National Competence Center in Bioinspired Materials, University of Fri\-bourg, Chemin des Verdiers 4, 1700 Fribourg, Switzerland\\
Email Address: bodo.wilts@plus.ac.at
\end{affiliations}


\begin{abstract}
While artificially photonic materials are typically highly ordered, photonic structures in many species of birds and insects do not possess a long-range order. Studying their order-disorder interplay sheds light on the origin of the photonic band gap. Here, we investigated the scale morphology of the \textit{Anoplophora graafi} longhorn beetle. Combining small-angle X-ray scattering and slice-and-view FIB-SEM tomography with molecular dynamics and optical simulations, we characterised the chitin sphere assemblies within blue and green \textit{A.\ graafi} scales. The low volume fraction of spheres and the number of their nearest neighbours are incompatible with any known close-packed sphere morphology. A short-range diamond lattice with long-range disorder best describes the sphere assembly, which will inspire the development of new colloid-based photonic materials.  
\end{abstract}

\section{Introduction}
Structural colour, ubiquitous in nature\cite{biroPhotonicNanoarchitecturesButterflies2011,srinivasarao_nano-optics_1999,sunStructuralColorationNature2013}, requires structural motives on the 100-nm length scale that give rise to the spectrally selective interference of light. A typical example is a stacked multilayer of different dielectric materials in which the Bragg criterion causes the reflection of light of a specific wavelength range, the so-called optical band gap \cite{joannopoulosPhotonicCrystalsMolding2008,stavenga_polarized_2011}. Although this is easily achieved for periodic structures, there are many examples in nature where partially disordered morphologies give rise to structural colour, either in the feathers of birds \cite{prumreview,yin_amorphous_2012} or in insects \cite{saranathan_structural_2015,10.1242/jeb.213306,bermudez-urenaStructuralDiversityVarying2020,djeghdi3DTomographicAnalysis2022,Parisotto2022,Bauernfeind2023 }. While some theories exist \cite{klattPhoamtonicDesignsYield2019,klattUniversalHiddenOrder2019,sellersLocalSelfuniformityPhotonic2017}, the detailed nature of the order-disorder interplay giving rise to structural colour is not well understood. Therefore, it is timely to study biological specimens in which evolution likely optimised this interplay \cite{saranathan_structural_2015,10.1242/jeb.213306,bermudez-urenaStructuralDiversityVarying2020,djeghdi3DTomographicAnalysis2022,Parisotto2022,Bauernfeind2023}. 

While structural analysis is relatively straightforward for ordered materials and can often be achieved from cross-sectional views \cite{wilts2016unique}, the characterisation of disordered morphologies is challenging and typically relies on the analysis of X-ray scattering data and multiple images of cross-sectional cuts. However, the ambiguity in analysing these data may often lead to wrong structural assignments of the underlying morphologies. To overcome these difficulties, very recently, 3D tomography was used to analyse natural photonic morphologies that lack long-range order \cite{djeghdi3DTomographicAnalysis2022, Parisotto2022, Bauernfeind2023}.    

One optical morphology found in nature is the seemingly random packing of 100-nm-sized spheres in birds \cite{Dufresne2009} and beetles \cite{dong_structural_2010,saranathan_structural_2015}. Studying \textit{Anoplophora graafi} (Ritsema, 1880), Dong and colleagues assigned the colour response of the beetle scales to the random close-packing of spheres, derived from cross-sectional images and numerical simulations \cite{dong_structural_2010}, while Saranathan \textit{et al}.\ termed the morphology ``quasi-ordered close-packed'' \cite{saranathan_structural_2015}. Although the terms ``random close-packing'' (RCP) and ``quasi-order'' have been put forward to justify the opening of a pseudo-bandgap, they still lack a rigorous definition \cite{torquatoRandomClosePacking2000}.

Being an age-old problem, packing spheres into a given volume is complex since it varies in how the packing is achieved. While ordered lattices have a well-defined number of nearest neighbours that touch each sphere, $n_\mathrm{touch}$, and well-defined volume fractions $\phi$, this is not the case for random packings. A characteristic case is the so-called maximally random jammed packing with $n_\mathrm{touch}\approx6$ and $\phi=0.64$. Structurally stable random packings (``strict jamming'') with the lowest volume fraction $\phi=0.60$ have more than six nearest neighbours ($n_\mathrm{touch}\approx6.3$) \cite{torquatoRandomClosePacking2000}. Apart from the particular case of ``tunnelled crystals'' \cite{torquato2007toward},  packings with $n_\mathrm{touch}<6$ and $\phi<0.6$ are not structurally stable.

In this work, we investigate the complete 3D structure in the elytral scales of an \textit{Anoplophora graafi} long\-horn beetle and relate it to the optical appearance of the scales. Scales were filled with Pt and 3D scale morphologies were reconstructed by slice-and-view tomography, using focused ion beam cutting combined with scanning electron microscopy imaging (FIB-SEM). This allowed statistical analysis of the sphere packing within the scales and the simulation of light propagation using finite-difference-time-domain (FDTD) optical simulations. 

\section{Results}
\subsection{General appearance and optical properties}
\begin{figure}[tbp]
    \centering
    {\includegraphics[width=\textwidth]{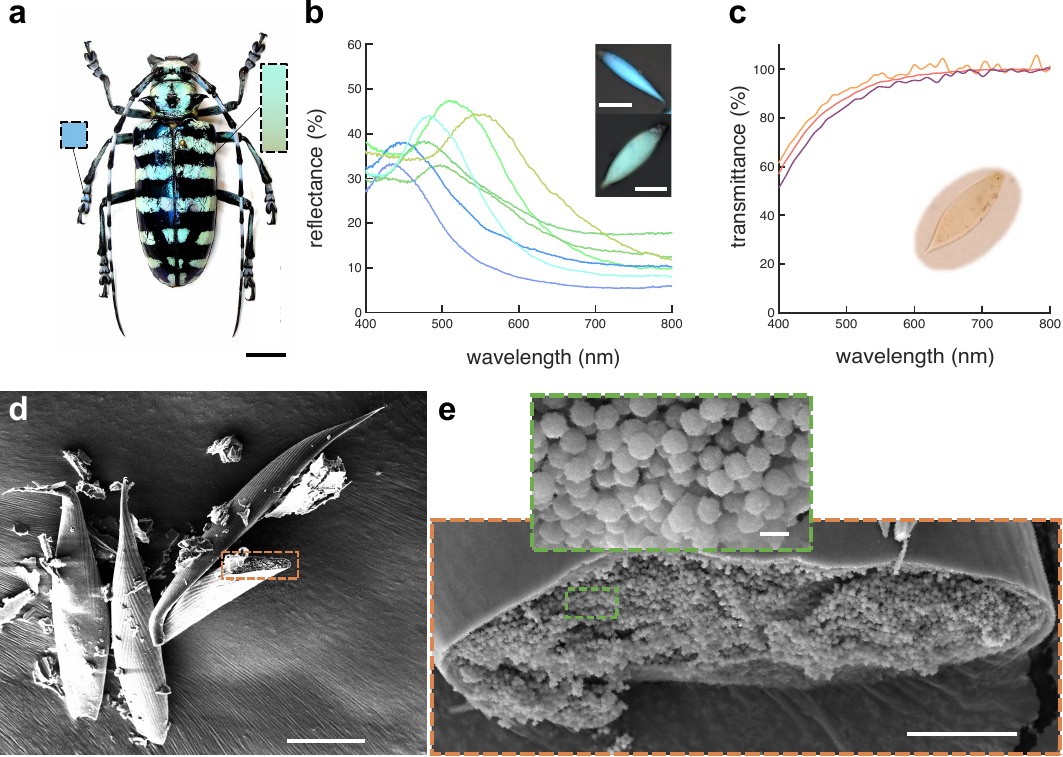}} 
    \caption[\textit{Anoplophora graafi} optical appearance and scale ultrastructure]{\textit{Anoplophora graafi} optical appearance and scale ultrastructure. (a) \emph{A. graafi} specimen with green and blue colouration on the main body and legs, respectively (photo credit: Vasiliy Feoktistov\,\textsuperscript{\textcopyright}). (b) Representative individual reflectance spectra from blue and green scales, calibrated using a silver mirror. The inset shows blue and green scales.  (c) Transmittance spectra of three different green scales immersed in index-matching oil ($n_\mathrm{o}=1.56$), showing short-wavelength absorption caused by a pigment. (d) SEM image of blue scales. with a broken scale marked and (e)  magnification of a different opened scale. The magnified inset shows chitin spheres packed within the scale. Scale bars: (a): 1\,cm; inset in (b): 50\,\textmu m; (d): 25\,\textmu m; (e): 5\,\textmu m; inset in (e): 500\,nm.}
    \label{fig:macromicrospectra}
\end{figure}

\textit{Anoplophora graafi} (Cerambycidae: Lamiinae; Ritsema, 1880) is an about 6\,cm long longhorn beetle native to Indonesia (Sumatra). It has black elytra with green horizontal stripes due to green-colored scales and legs that are covered with blue scales that are typically thinner and more elongated than the green scales (Fig.\,\ref{fig:macromicrospectra}a,b,d). 

Micro-spectrophotometry was performed on green and blue scales scratched off the beetle elytra and legs, respectively. Each scale gave rise to a single reflection peak in the blue-yellow spectral range with a small increase towards 400\,nm. Several representative reflectance spectra are shown in Figs.\,\ref{fig:macromicrospectra}b and\,\ref{fig:diam_spectra}. The blue scales typically display peak reflectance of up to 40\% (against a silver mirror reference) in the 430--520\,nm wavelength range. The green scales are more strongly reflective, and their peaks are bathochromically shifted, with maximal reflectance of up to 50\% in the 530--590\,nm wavelength range. Combined, the scales cover a wavelength range from blue to yellow. Note that some pale-red scales were also found ($\lambda_\mathrm{max}=635$\,nm, Fig.\,\ref{fig:diam_spectra}, c.f. ref.\,\cite{dong_structural_2010}). 

The presence of pigments within the scales was investigated by immersing the scales in oil ($n=1.56$) that matches the refractive index of chitin \cite{leertouwer_refractive_2011} to suppress the contribution of the scale ultrastructure. In transmission, the immersed scales appear pale yellow (Fig.\,\ref{fig:macromicrospectra}c, inset) and the transmittance spectra of Fig.\,\ref{fig:macromicrospectra}c show a continuing decrease from about 550\,nm towards shorter wavelengths. We attribute this decrease to pigment absorption, probably caused by melanin \cite{stavenga_polarized_2011}. A comparison of the reflection and (index-matched) transmittance spectra of Fig.\,\ref{fig:macromicrospectra}b,c indicates that a photonic structure determines the colour appearance of the scales.

\subsection{Ultrastructure of the wing scales}
To image the ultrastructure of the blue and green scales that gives rise to colouration, we used scanning electron microscopy (SEM). Scales broken during sample preparation allowed for the observation of their internal structure (Fig.\,\ref{fig:macromicrospectra}d,e). The scales are hollow elongated sack-like, surrounded by an $\approx$ 500\,nm thick cortex. The hollow cavity of the scales (Fig.\,\ref{fig:macromicrospectra}e) is filled with chitin spheres that appear uniform in size. Analysing the spheres present in blue and green scales from the SEM images revealed diameters covering a range between 200 and 295\,nm (see Fig.\,S\ref{fig:diam_spectra}). 

\begin{figure}[tbp]
    \centering
    {\includegraphics[width=0.85\linewidth]{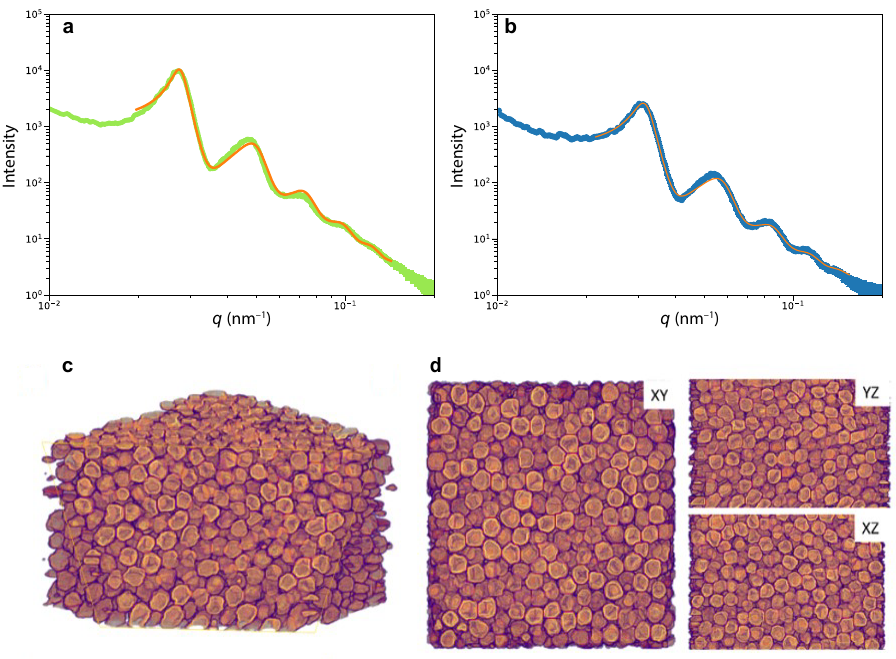}} 
    \caption{Ultrastructure of the scales. 
    (a,b) SAXS spectra of (a) green and (b) blue scales. The plots were obtained from the radial integration of 2D SAXS patterns. The orange lines are fits of an assembly of hard spheres to the data. 
    (c,d) Tomogram of a green scale. (c) 3D view of a reconstructed experimental volume obtained by FIB-SEM slice-and-view tomography, and (d) $XY$, $YZ$, and $XZ$ cross-sectional views. The imaged volume was $V=2805\times3015\times1905\,$nm$^3$.}
    \label{fig:structure}
\end{figure}

To more accurately and statistically measure the particle diameter, we performed small-angle X-ray scattering (SAXS). Fitting the SAXS spectra of Fig.\ \ref{fig:structure}a (see Supplementary Information, SI, for details) of green and blue scales yielded sphere diameters of $d_\mathrm{g}=252\pm14$\,nm and $d_\mathrm{b}=216\pm13$\,nm, and volume fractions of $\phi_\mathrm{g}\approx51$\% and $\phi_\mathrm{b}\approx40$\%, respectively.  The peak sequence in SAXS spectra reveals short-range positional correlations that decay over distances of about 4 sphere diameters.    
 
More detailed insight into the arrangement of the chitin spheres within the scales requires real-space 3D imaging. To this end, the cortex surrounding a green scale was removed using Ar plasma etching. In the chamber of a scanning electron microscope equipped with a Ga focused ion beam (FIB-SEM), the porous volume of the scale was filled with platinum by electron beam-induced deposition (Pt-EBID), followed by a slice-and-view FIB-SEM tomography (see Experimental Section for details, also ref.\,\cite{djeghdi3DTomographicAnalysis2022}). Detailed views of the resulting volumetric reconstruction are shown in Fig.\,\ref{fig:structure}c,d.
The tomogram yielded a volume $V=2.8\times3.0\times1.9$ \textmu m$^{3}$ filled with deformed spheres of uniform diameter of $d_\mathrm{tomo}=208 \pm 16$\,nm and a volume fraction $\phi_\mathrm{tomo}=51\%$. 

The diameter discrepancy between the SAXS analysis and tomographic imaging probably arises from the Pt-EBID protocol, where the samples are exposed to high electron doses, resulting in a shrinkage of $\approx10-20$\% (see also \cite{Bauernfeind2023}). Furthermore, the spheres in the tomogram in Fig.\ \ref{fig:structure}c,d appear more strongly deformed compared to the SEM images of Fig.\ \ref{fig:macromicrospectra}e, probably due to the combined effect of beam damage and reconstruction artifacts that occur when creating a 3D representation from 2D FIB-SEM slices.  

\begin{figure}[tbp]
    \centering
    {\includegraphics[width=.8\linewidth]{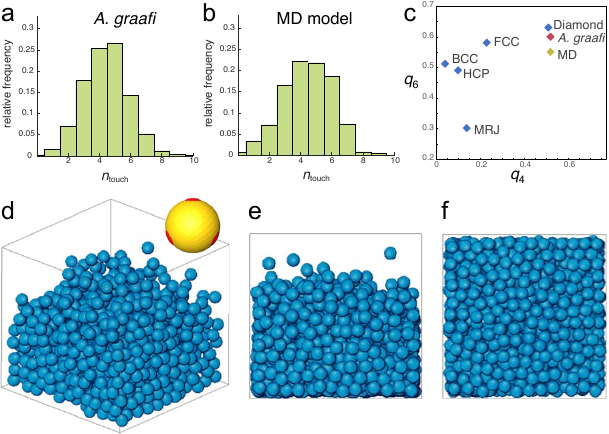}} 
    \caption{Nearest neighbour distibutions and molecular dynamics simulation. (a,b) Number of touching neighbours $n_\mathrm{touch}$ determined from the tomogram in Fig.\ \ref{fig:structure}c,d and the sphere assembly in (d-f), respectively. (c) Plotting the averaged bond-orientational order parameters\cite{steinhardtBondorientationalOrderLiquids1983} $q_6$ vs $q_4$ allows identifying close-packed motifs.  The calculated values for the tomogram of the \textit{A.\ graafi} scale (Fig.\ \ref{fig:structure}c,d) and the MD simulation (Fig.\ \ref{fig:md}a) are compared to HCP, BCC, FCC, diamond reference structures\cite{10.3389/fmats.2017.00034}, and to a maximally random, jammed (MRJ) sphere assembly \cite{PhysRevLett.109.205701}. (d-f) Molecular dynamics simulation ($N\!=\!1080$) of spheres with tetragonally arranged attractive patches (inset in (d); attractive well depth of $8k_\mathrm{B}T$). All other sphere-sphere contacts are repulsive.  A weak gravitational force was adjusted such that the sphere assembly equilibrated to a volume fraction of $\phi_\mathrm{MD}=46\%$. (d) 3D representation, (e) side view, and (f) bottom view.}
    \label{fig:md}
\end{figure}

Nevertheless, the tomography dataset allows a statistical analysis of the sphere assembly. Allowing for a 15\,nm gap, accounting for the finite voxel size of approx.\ 15\,nm, the number of touching spheres ($n_\mathrm{touch}$) was evaluated for all spheres of the tomography dataset. The histogram in Fig.\ \ref{fig:md}a peaks at $n_\mathrm{touch}=4.5\pm2$.

\subsection{MD simulations of particle assemblies and order parameters}
Since random closed-packed sphere assemblies with $n_\mathrm{touch}<6$ are not structurally stable, we performed molecular dynamics (MD) simulations assuming ``sticky'' particles, where each sphere has 4 tetragonally arranged ``glue'' patches (Fig.\ \ref{fig:md}d, inset). To correspond to the SAXS values for the blue scale, we chose particles with 210 nm diameter. The patches attract each other with a short-range Lennard–Jones potential with a well-depth of $8k_\mathrm{B}T$ (where $k_\mathrm{B}$ is Boltzmann's constant and $T$ is the temperature), while all other possible contacts are repulsive. The density of the particle assembly was controlled by introducing a variable ``gravity force'' (see Experimental Section for details).     

Sufficiently strong gravity forces led to an ordered cubic arrangement of the sphere assembly, with $n_\mathrm{touch}=8$, as expected for a body-centred-cubic arrangement of repelling hard spheres. Above a threshold in the strength of the imposed gravity field, a distribution of $n_\mathrm{touch}$ values centred between 4 and 5 was obtained. In this regime, the volume fraction varied with the gravitational potential. An MD simulation best matching the experimentally determined parameters is shown in Fig.\ \ref{fig:md}d--f ($N=1080$, $\phi_\mathrm{MD} = 0.46$). The histogram in Fig.\,\ref{fig:md}b closely matches that of the experimentally determined histogram of Fig.\ \ref{fig:md}a, with $n_\mathrm{touch}=4.5\pm2.5$.

A quantitative measure allowing to characterise the local symmetry of a sphere assembly was introduced by Steinhardt and coworkers \cite{steinhardtBondorientationalOrderLiquids1983}. Drawing bonds between each particle of an assembly and their nearest neighbours defines the angles between these bonds. The bond-orientational order parameters $q_l$ are then defined as lowest-order rotation-invariant of the $l$\textsuperscript{th} moment in a multipole expansion of the bond vector distribution on a unit sphere \cite{10.1063/1.4774084} (Experimental Section).  The fourth- and sixth-order parameters $q_4$ and $q_6$ are often used to quantify the local order of a sphere assembly, e.g., how closely they resemble hexagonal close-packed (HCP), body-centred cubic (BCC), or face-centred cubic (FCC) structures. Figure \ref{fig:md}c shows a plot of $q_6$ and $q_4$ for these reference structures alongside with the values computed from the tomogram in Fig.\ \ref{fig:structure}c,d and the MD simulation in Fig.\ \ref{fig:md}d--f.  These values differ not only from the reference symmetries but also from a maximally random jammed sphere assembly (MRJ) \cite{PhysRevLett.109.205701}. The $q_4$-$q_6$ order parameter of a diamond lattice\cite{10.3389/fmats.2017.00034      } closely matches those of the tomogram and MD simulation.

\begin{figure}[tbp]
    \centering
    {\includegraphics[width=\linewidth]{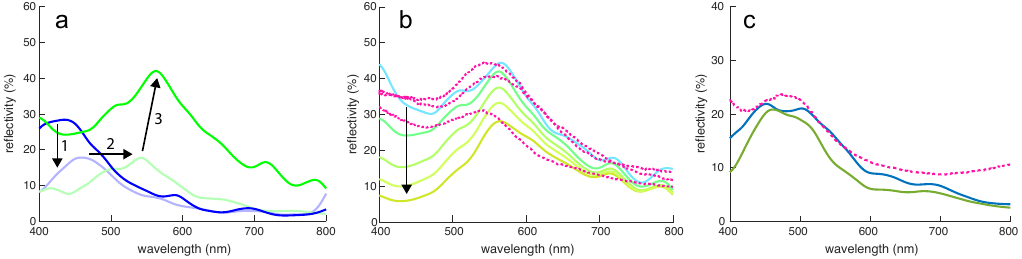}} 
    \caption{FDTD simulations. (a) Simulated reflection spectra of the tomogram in Fig.\ \ref{fig:structure}c,d with a sphere refractive index of $n=1.56$ (chitin), without (dark blue) and with an added melanin absorber. Since the electron irradiation during (Pt-EBID) causes a shrinkage of the sample, a 20\% scaled-up sample of melanized chitin spheres is also shown (light green). The deformed tomography spheres were replaced by perfectly round spheres (dark green line), increasing the main reflection peak to about 40\%. 
    (b) Influence of adding increasing amounts of pigment (indicated by the arrow) on the reflectivity, compared to three experimental spectra (dashed lines). 
    (c) Simulated reflection spectra for the MD simulation shown in Fig.\ref{fig:md}d--f (blue curve) compared to an experimental spectrum obtained from a blue scale (dashed line). The green line shows the reflectance of an ordered sphere assembly obtained for a strong gravitational field in the MD simulation.
    }
    \label{fig:FDTD}
\end{figure}

\subsection{Optical modelling}
The optical properties of the digital representations of the tomograms and the MD simulations were explored by finite-difference-time-domain (FDTD) simulations. 

The tomogram of Fig.\ \ref{fig:structure}c, comprising spheres of $\approx210$nm in diameter and a volume fraction of $\approx 0.5$ were used as input for FDTD calculations to simulate the reflection spectra of this sphere assembly, assuming the refractive index of chitin ($n=1.56$; ref.~\cite{leertouwer_refractive_2011}). Figure \ref{fig:FDTD}a shows a reflection peak at $\lambda=430$\,nm with a peak reflectance of 28\%, 
indicative of structural colouration. Since Fig.\ \ref{fig:macromicrospectra}c indicates the presence of a broadband-absorbing pigment, an imaginary refractive index was included in the simulation to include the effects of melanin (sensu\,\cite{stavenga_sexual_2012}). This strongly reduced the simulated reflectivity around 400\,nm, mostly preserving the 430\,nm reflection peak. 

The experimental spectra of green scales in Fig.\ \ref{fig:macromicrospectra}b show reflection maxima in the green-yellow region, much above the $\lambda=430$\,nm FDTD result. As described above, the Pt-EBID protocol led to a volume shrinkage of $\approx20$\%. In further simulations, the voxel volume of the tomogram was therefore scaled by 20\% to compensate for this shrinkage. This caused a red-shift of the spectra, moving the peak at 430\,nm to 540\,nm in Fig.\,\ref{fig:FDTD}a, in the range of the experimental results. Finally, since beam damage combined with 3D imaging artefacts caused the deformation of the reconstructed chitin spheres, a 3D dataset was generated, in which each of the particles was replaced with a perfect sphere of diameter 252\,nm, the value obtained from the SAXS measurement of Fig.\ \ref{fig:structure}a. The resulting reflection spectrum in Fig.\,\ref{fig:FDTD}a (green line) exhibited an increased reflectivity of ca.\ 40\%  at $\lambda=566$\,nm. While the pronounced non-spherical shape of the spheres in Fig.\ \ref{fig:structure}c,d is likely caused by beam damage and the 3D reconstruction algorithm (see above), a close examination of the spheres in Fig.\ \ref{fig:macromicrospectra}e reveals a slightly rough surface structure of the chitin spheres. The FDTD simulations of assemblies of rough spheres, shown in Fig.\ \ref{fig:roughness}, indicate a negligible effect of sphere surface roughness, as long as the sphere volume is conserved.

Figure \ref{fig:structure} therefore illustrates that three transformations of the initial (transparent) tomogram are required to match the experimental results: (1) the suppression of UV reflectivity by pigment addition, (2) the volume upscaling to compensate for e-beam shrinkage, and (3) the replacement of deformed lower volume spheres by larger volume spheres with a low surface roughness (see also Fig.\ \ref{fig:roughness}).

The influence of pigment addition was further investigated in Fig.\ \ref{fig:FDTD}b for the assembly of 252\,nm spheres of Fig.\ \ref{fig:FDTD}a. Starting from the transparent case of pure chitin (light blue line), the magnitude of the imaginary part of the refractive index was successively increased, indicated by the arrow in Fig.\ \ref{fig:FDTD}b. The used refractive index dispersions are shown in Fig.\ \ref{fig:RI}. A good match with experimental spectra was reached for moderate values of the absorptivities. Figure \ref{fig:FDTD}c shows an experimental reflection spectrum of a blue scale compared with the FDTD simulations of the MD sphere assemblies of Fig.\ref{fig:md}d--f, showing a good match. An FDTD spectrum of a more ordered simulated system is also shown, displaying a quite similar spectral response with less back-scattering at small wavelengths. In all three cases, the main reflection peak is around 450--480\,nm with a reflectivity of 20-25\%.

\section{Discussion}
The disordered colloidal assembly present in the scale of an \textit{Anoplophora graafi} longhorn beetle is able to generate a range of colours from light blue to green-yellow by only modifying the diameter of the chitin spheres aggregated within the scale (Fig.\,\ref{fig:diam_spectra}). Longhorn beetles are known to generate the vivid colouration of their scales using structural colouration, and both, well-ordered\cite{kobayashi_discovery_2021,colomer_photonic_2012,liu_structural_2009, bermudez-urenaStructuralDiversityVarying2020} and disordered systems have been previously described \cite{dong_structural_2010,dong_optical_2011, bermudez-urenaStructuralDiversityVarying2020, Bauernfeind2023}.

The SAXS data analysis reveals sphere sizes, within the error margin, of approximately one-half of the peak wavelength of the reflection maxima of the blue and green scales. The decaying oscillation of the spectra is typical for local packing correlations decaying over 4--5 sphere diameters. Surprisingly, the volume fractions are much below those for stable random packed sphere assemblies of $\phi>0.6$.

The 3D tomography data provide further insight into the sphere assembly. Since the sample exhibited relatively substantial volume reduction caused by the Pt-EBID protocol \cite{djeghdi3DTomographicAnalysis2022,Bauernfeind2023}, two corrections were required before further analysing the data, (1) the isotropic expansion of the tomogram by a factor of 1.2, which was achieved by rescaling the voxel volume and (2) smoothing of the chitin spheres by replacing them with ideal spheres. Analysing the coordination number for each sphere in the tomogram, a distribution with a maximum of $n_\mathrm{touch}\approx 4.5$ was determined. This $n_\mathrm{touch}$ value is far below that of typical random close-packed assemblies with $n_\mathrm{touch}\approx 6$, close to $n_\mathrm{touch}=4$ of a diamond lattice.  

Local lattice similarities can be quantified by the averaged bond-orientational order parameters $q_4$ and $q_6$ \cite{steinhardtBondorientationalOrderLiquids1983} (see Experimental Section). In a graph of $q_6$ vs $q_4$ in Fig.\ \ref{fig:md}c, a close match of the parameters determined from the tomogram and a diamond lattice are found, well separated from the other closed-packed morphologies.

Dong \cite{dong_structural_2010} and others \cite{saranathan_structural_2015} have previously described the structure in this beetle as quasi-ordered. The unusually low volume fractions of the chitin spheres revealed by our SAXS and FIB-SEM tomography suggest that structures within the \textit{A. graafi} scales are not a close-packed sphere morphology but have a local morphology of a diamond lattice, where four nearest neighbours surround each sphere.  

The MD simulations of Fig.\ \ref{fig:md}d-f were motivated by the experimentally found value of $n_\mathrm{touch}\approx 4$, and the result of Fig.\ \ref{fig:md}c. By using colloids with attractive patches arranged in a tetrahedral pattern (and repulsive elsewhere), MD simulations in the presence of a weak gravitational potential produced a sphere assembly that mirrored that of the tomography results of Fig.\ \ref{fig:structure}b,c. The distribution of $n_\mathrm{touch}$-values in Fig.\ \ref{fig:md}b closely mirrors the experimental histogram in Fig.\ \ref{fig:md}a, as do the $q_4$, $q_6$ order parameters in Fig.\ \ref{fig:md}c. The MD simulations suggest that low-sphere volume fractions are enabled by specific sphere adhesion (``glue''), allowing the formation of a tetrahedrally coordinated sphere assembly with volume fractions below the lower limit for random close packings.

While the local morphology of the sphere assembly is relatively regular, with sphere polydispersities on the order of 10\% and narrow $n_\mathrm{touch}$-distributions, long-range order is absent. Structurally, this arises from the absence of correlations in the bond-rotation angle between neighbouring tetrahedrally connected nodes \cite{djeghdi3DTomographicAnalysis2022}. The described locally tetrahedrally correlated but long-range disordered morphology is reminiscent of a recent theoretical study \cite{sellersLocalSelfuniformityPhotonic2017}. Starting from a periodic, tetrahedrally interconnected strut network disorder is introduced through different bond-angle distributions. This study shows that the optical band gap of the ordered lattice is preserved in a range of bond-angle permutations. Note also that networks constructed in this fashion suppress density fluctuations on long length scales and are therefore hyperuniform \cite{florescuDesignerDisorderedMaterials2009}. Unfortunately, little is known about the growth of these structures in vivo, and it will be interesting to investigate this further, given the large variability of ultrastructures in beetles \cite{saranathan_structural_2015,bermudez-urenaStructuralDiversityVarying2020,Bauernfeind2023}.

\section{Conclusion}
Combining SAXS and FIB-SEM structural measurements, MD simulations with optical measurements, and FDTD simulations, the interplay between the 100-nm morphology within \textit{A. graafi} scales and their optical response were studied. Despite the apparent lack of order in the assembly of chitin spheres, the chitin sphere assemblies are locally relatively well-defined, closely resembling a diamond morphology with $n_\mathrm{touch}$-values centred around 4.5. Particularly, comparing the fourth- and sixth-order bond-orientational parameters with various ordered and disordered sphere assemblies yields a close match with the diamond lattice. A molecular dynamics simulation of an assembly of tetrahedral patchy particles also provided a close match in terms of structural and optical parameters. This study, therefore, paves the way towards the manufacture of disordered photonic materials with isotropic bandgaps in the visible wavelength range.

\section{Experimental Section}

\subsection{Optical characterisation}
The spectral characterisation of the scales was performed using a xenon light source (Thorlabs SLS401; Thorlabs GmbH, Dachau, Germany) and a ZEISS Axio Scope.A1 microscope (Zeiss AG, Oberkochen, Germany). The light reflected from the sample was collected by an optical fibre (230\,\textmu m core) in a plane confocal to the image plane, resulting in an effective measurement spot diameter of $\sim$13\,\textmu m at a magnification of $\times20$. A spectrometer (Ocean Optics Maya2000 Pro; Ocean Optics, Dunedin, FL, USA) was used to record the spectra. Optical micrographs were captured with a CCD camera (GS3-U3-28S5C-C, FLIR Integrated Imaging Solutions Inc., Richmond, Canada). Reflection spectra were taken on scales still attached to the elytron, where a protected silver mirror (PF10-03-P01, Thorlabs) served as a reference standard. For transmission spectra, the scales were detached from the elytron and deposited on a microscope glass slide before immersion in a refractive index matching oil ($n_\mathrm{o}=1.56$).

\subsection{Electron imaging and FIB-SEM tomography}
Individual scales were gently scratched from the elytron and deposited on an aluminium SEM stub (Plano-EM, Wetzlar, Germany) covered by conductive carbon tape. The cortex layer surrounding individual scales was partially removed by plasma etching using a PE-100-RIE system (Plasma Etch Inc., Nevada, USA). The samples were exposed to a 4:20 \ce{O2}/Ar plasma mixture for 12 to 18 minutes. The stub was then sputter coated with a 7\,nm thick layer of either gold or platinum using a Cressington 208 HR (Cressington Scientific Instruments, Watford, England) sputter coater. Copper tape and silver paste were also added to increase the conductivity of the sample and limit charging effects. Top-view scanning electron microscopy (SEM) pictures were taken using a Tescan Mira3 (Tescan, Brno, Czechia) with a beam voltage of 8\,kV and a working distance of 10\,mm. Cross-sections were milled and imaged using a Thermo Scientific Scios 2 DualBeam FIB-SEM (FEI, Eindhoven, the Netherlands).

To enable 3D reconstruction, the scales were first filled \textit{in situ} with platinum by electron beam-induced deposition (Pt-EBID) \cite{eswara-moorthy_situ_2014, djeghdi3DTomographicAnalysis2022}. A gaseous precursor, \ce{C9H16Pt}, was injected near the surface of the cortex-free region of interest (ROI) through the gas-injection system needle and dissociated into Pt by interacting with the electron beam, set to an acceleration voltage of 30\,kV, a current of 1.6\,nA and a dwell time of 15\,\textmu s. These parameters were optimised to enable complete infiltration throughout the entire scale thickness. A rough cut was milled in front of the ROI to expose the imaging plane, and trenches were milled on the sides to provide deposition sites for the milled material, preventing redeposition onto the imaged section. A fiducial was created to correct the image stack for beam-induced drift, stage displacement, and tilt. The tomography process was automatized using ThermoFisher ASV software (v.\ 4). 15\,nm thick slices were milled with the \ce{Ga+} beam set to an  of 30\,kV and a beam current of 0.30\,nA. Images were acquired after each slice in the OptiTilt configuration using the built-in SEM Everhart-Thornley (ETD, secondary electrons) and in-lens T1 (A+B composite mode, back-scattered electrons) detectors. The electron beam was set to a voltage of 2\,kV. The built-in tilt correction feature was used to compensate for the image distortion induced by the acquisition at a 52\textdegree\ angle.

\subsection{3D reconstruction}
Fiji \cite{schindelin_fiji_2012} and FEI Avizo\textsuperscript{\texttrademark} for Materials Science 2020.2 software were used for processing of image stacks, 3D reconstruction, and statistical analysis. First, a registration of the images in the stack was performed using a combination of the Fiji \textit{StackReg} \cite{thevenaz_pyramid_1998} and \textit{Correct 3D Drift} plug-ins \cite{parslow_sample_2014}. Subsequently, a median filter was applied to despeckle and smooth the images, which were then inverted to make the chitinous material appear white and the Pt filling (corresponding to the air network) black. Segmentation was done using the trainable Weka segmentation 3D plug-in \cite{arganda-carreras_trainable_2017} in Fiji. The ImageJ watershed algorithm was applied to separate the spheres into individual particles \cite{Vincent/Soille:1991}. Separate objects can then be individually labelled, and their position and shape descriptors can be extracted using Avizo.

 \subsection{FDTD simulations}
Finite-Difference Time-Domain (FDTD) calculations were performed using the Lumerical software (Ansys Lumerical FDTD 2021; Ansys Inc, Canonsburg, PA, USA) to simulate reflection spectra and photonic density of states (PDOS), directly using the obtained 3D reconstructions (tomograms or MD simulations) as input. Simulations were performed in 3D with PML absorbing boundary conditions along the incidence direction of light and periodic boundary conditions in the other two directions. The volume was illuminated with a broadband light source at normal incidence with wavelengths 300--800\,nm.

A complex refractive index that varied with the wavelength was chosen to model the optical properties of chitin and the absorbing melanin-like pigment in the scales (based on \cite{stavenga_sexual_2012}, see Fig.\ \ref{fig:RI}). Since the pigment contribution is mainly absorptive, the real part of the refractive index was set to that of chitin, which is well approximated by a wavelength-dependent Cauchy law \cite{leertouwer_refractive_2011}. 

\subsection{Small-Angle X-ray Scattering (SAXS)}
Ultra-Small-Angle X-ray Scattering (USAXS) was performed on the ID02 beamline at the European Synchrotron Radiation Facility (ESRF) in Grenoble (France) \cite{narayanan2022performance}. The X-ray wavelength was 0.1\,nm and the sample-to-detector distance was fixed at 31\,m to cover a q-range of about $\approx 0.004-0.2\,\mathrm{nm}^{-1}$, where \textit{q} is the scattering vector. At the sample position, the photon flux was about 1010 photons s$^{-1}$, and the beam size was 120\,\textmu m $\times$ 63\,\textmu m (horizontal and vertical dimensions). Scattering patterns were recorded on a flight tube-enclosed EIGER2 4M detector (Dectris) with a 0.1\,s exposure time. The samples were sufficiently translated after exposure to avoid beam damage by X-ray overexposure. Several individual scales were scratched off the insect and sandwiched between two pieces of Kapton tape. Radial integration of 2D SAXS patterns was performed to obtain 1D plots. The fits were obtained with the software SasView\footnote{http://www.sasview.org/} 5.0.5 using model spheres with a Guinier form factor and an interparticle structure factor for polydisperse spherical particles interacting through hard sphere (excluded volume) interactions, employing  the Percus-Yevick closure relationship. These fits yield a Gaussian distribution of the sphere diameters and their volume fraction.

\subsection{Calculation of the order parameter}

The bond orientational parameters were computed using the Python pyboo package \cite{pyboo}, which implements a theory developed by Steinhardt  \cite{steinhardtBondorientationalOrderLiquids1983} for the classification of crystal structures \cite{lechnerAccurateDeterminationCrystal2008}. In short, the local tensorial bond orientational order parameter $q_{lm}$ and its second-order invariant $q_l$ are defined as
\begin{equation}
\begin{cases}q_{lm}\left( i\right) =\dfrac{1}{N_{i}}\sum ^{N_i}_{0}Y_{lm}\left( \theta _{ij},\phi _{ij}\right) \\ q_{l}=\dfrac{4\pi }{2l+1}\sum ^{l}_{m=-l}\left| q_{lm}\right| ^{2}\end{cases},
\end{equation}
where $N_i$ bonds start from position $i$. The spherical harmonics $Y_{lm}$ and the Legendre polynomials $P_l^m$ are defined as
\begin{equation}
\begin{cases} Y_{l}^{m}\left( \theta,\phi\right) =\left( -1\right) ^{m} \sqrt{\dfrac{2l+1}{4\pi }\dfrac{\left( l-m\right) !}{\left( l+m\right) !}} P_{l}^{m}\left( \cos \left( \theta\right) \right) e^{im\phi } \\
P_{l}^{m}\left( x\right) =\dfrac{1}{2^ll!}\left( 1-x^{2}\right) ^{\dfrac{m}{2}}\dfrac{\partial ^{l+m}}{\partial x^{l+m}}{\left(x^{2}-1\right) ^{l}}. \end{cases}.
\end{equation}


Note that these parameters are highly dependent on the definition of the neighbourhood of a particle. Similar to the calculation of the coordination number, a particle $i$ is considered a neighbour of a particle $j$ if the distance $d_{ij}\leq r_i + r_j + 15$\,nm, where 15\,nm corresponds to the voxel side length.

\subsection{HOOMD-blue simulations}
The chitinous spheres were modelled as patchy particles as described in \cite{heColloidalDiamond2020,ANDERSON2020109363}. In short, the simulations were performed using a short-range attractive Lennard–Jones potential $U_p$ between patches, given by $U_p(r) = 4\epsilon[(\sigma/r)2n-(\sigma/r)n]$ with $n = 24$, where $r$ is the distance between the centres of the spheres and $\sigma$ is the radius of the spheres. This places the minimum in the potential at $r = 2^{1/n}\sigma = 1.03\sigma$. The well depth $\epsilon$ sets the energy scale (in units of $k_\mathrm{B}T$, where $k_\mathrm{B}$ is Boltzmann's constant and $T$ is the temperature), which we arbitrarily set to 8 for the attractive interaction between the patches. An additional weak attractive interaction between the main bodies was modelled by a Lennard–Jones potential with $n = 24$ and $\epsilon = 3$. The interaction between the main body and the patches was modelled by a short-range repulsive Weeks–Chandler–Andersen (WCA) potential $U_\mathrm{c}(r) = 4\epsilon[(\sigma/r)2n - (\sigma/r)n + 1/4]$ for $r \leq 2^{1/n}\sigma$ and $U_\mathrm{c}(r) = 0$ for $r > 2^{1/n}\sigma$, with $n = 24$ and $\epsilon = 10$. In a typical simulation run, 1080 particles were placed in a periodic box. The final system configuration was extracted using the Ovito software and was further analysed using the Avizo/Amira package.

\medskip
\textbf{Supporting Information} \par 
Supporting Information is available from the Wiley Online Library or from the author.




\medskip
\textbf{Acknowledgements} \par 
This study was supported by a European Research Council (ERC) Advanced grant (PrISMoID, 833895), the National Center of Competence in Research Bio-Inspired Materials of the Swiss National Science Foundation (SNSF) and the Adolphe Merkle Foundation. We acknowledge the help of John Gale with the MD simulations and useful discussions with Viola Bauernfeind.  

\medskip
\textbf{Conflicts of Interest} \par 
The authors declare no conflict of interest.

\medskip

%
\bibliographystyle{MSP}
\bibliography{biblio_agraafi}

\clearpage
\setcounter{figure}{0}
\renewcommand{\thefigure}{S\arabic{figure}}
\fancyhead{} 
\section*{Supplementary Information}
\begin{figure}[h!]
    \centering
    {\includegraphics[width=0.85\linewidth]{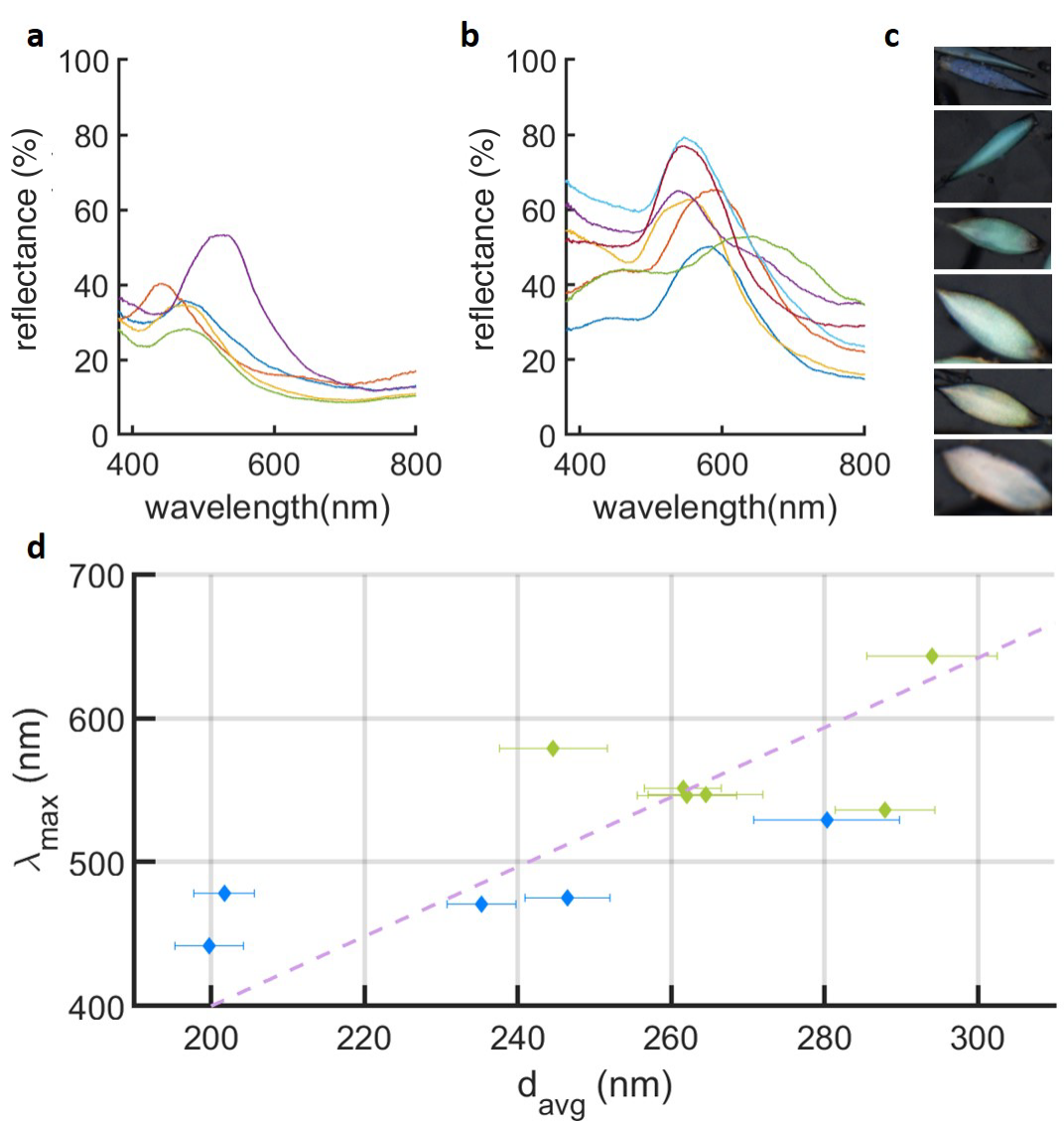}} 
    \caption[Relationship between sphere diameter and reflected wavelength]{Relationship between sphere diameter and reflected wavelength. (a) Reflectance spectra of scales collected from legs and (b) the main body, measured against a white diffusive reference tile. (c) Examples of scales covering the entire visible wavelength range. (d) The diameter is an average from SEM image analyses similar to that of Fig.\ \ref{fig:macromicrospectra}e ($N$=15). The regression line is a guide to the eye. 
    }
    \label{fig:diam_spectra}
\end{figure}

\begin{figure}[tbp]
    \centering
    {\includegraphics[width=0.85\textwidth]{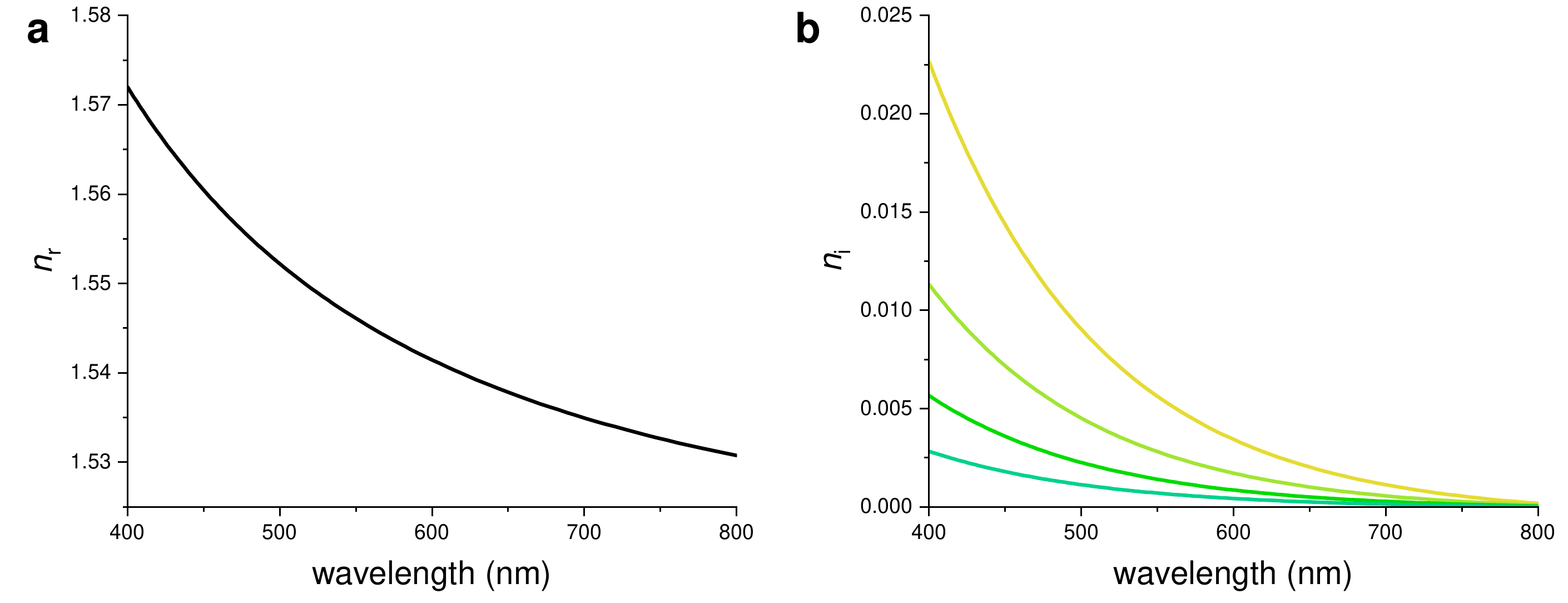}} 
    \caption{Refractive index dispersion as used in the FDTD modelling of Fig.\ \ref{fig:FDTD}. Due to the relatively small amount of pigment, the real part of the refractive index was assumed to be identical to cuticular chitin \cite{leertouwer_refractive_2011} and the imaginary part was varied. Color scheme of the curves is identical as in Fig.\ \ref{fig:FDTD}.}
    \label{fig:RI}
\end{figure}

\begin{figure}[tbp]
    \centering
    {\includegraphics[width=0.85\textwidth]{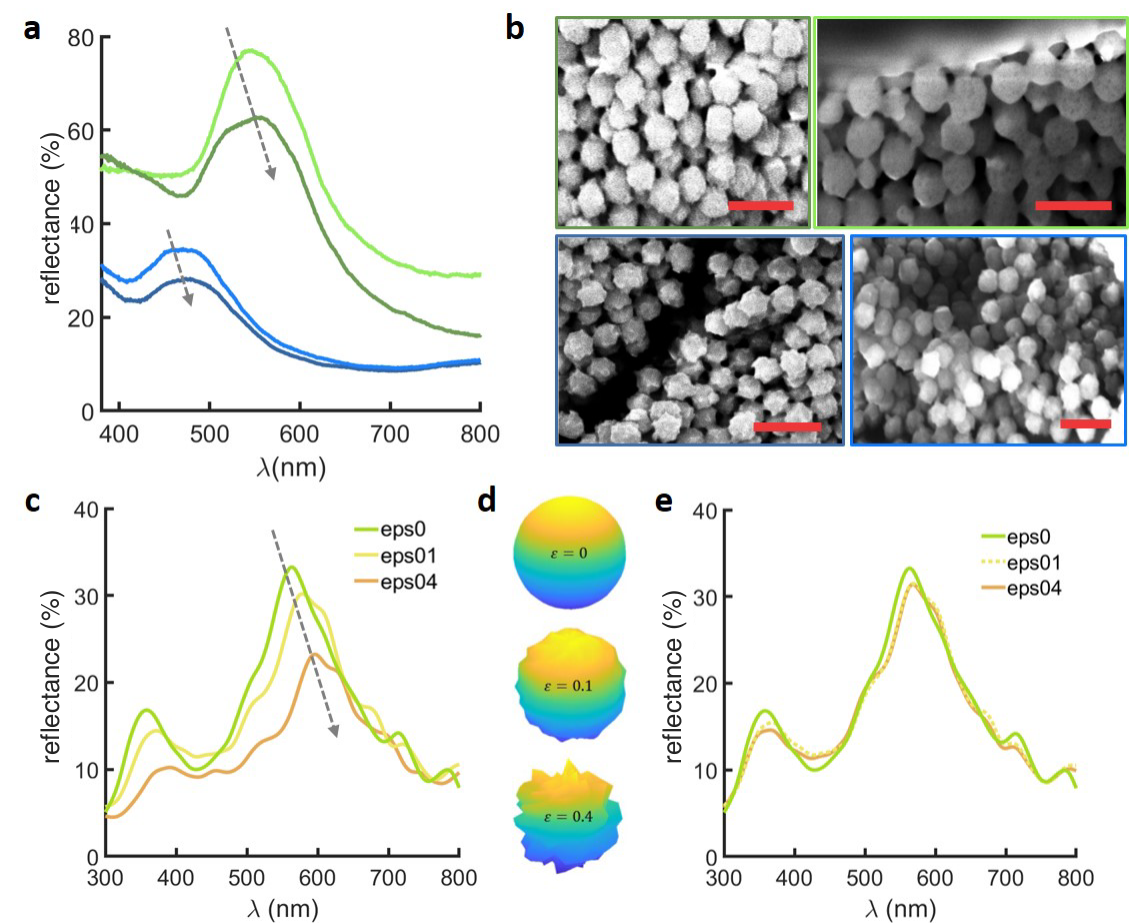}} 
    \caption{Effect of sphere roughness. (a) Reflectance spectra from two green and two blue scales, shown in (b), containing spheres with differing surface roughness. The spectra of the rougher spheres are slightly red-shifted with a lower overall reflectance (green scale: $\Delta\lambda=14$\,nm, $\Delta R=14$\%; blue scale: $\Delta\lambda=4$\,nm,  $\Delta R=7$\%). The scale bars in (b) are 500\,nm. (c) FDTD simulated reflectivity spectra for the tomogram of Fig.\ \ref{fig:structure}b,c, replacing the measured objects with perfectly smooth spheres and adding increasing surface roughness, as indicated in (d). The increasing spectral red-shift with increasing surface roughness arises from the corresponding decrease in sphere volume. When scaling the spheres to the same volumes in (e), keeping the overall filling volume fraction equal, the spectra of smooth and rough spheres overlap.}
    \label{fig:roughness}
\end{figure}
\end{document}